\documentclass[manuscript, acmconf, authorversion]{acmart}
\usepackage{epigraph} 
\usepackage{algorithm}
\usepackage{algpseudocode}
\usepackage{caption}
\usepackage{subcaption}

\AtBeginDocument{%
  \providecommand\BibTeX{{%
    \normalfont B\kern-0.5em{\scshape i\kern-0.25em b}\kern-0.8em\TeX}}}

\setcopyright{acmlicensed}\acmConference[ARTECH 2023]{11th International Conference on Digital and Interactive Arts}{November 28--30, 2023}{Faro, Portugal}
\copyrightyear{2023}
\acmYear{2023}
\acmDOI{10.1145/3632776.3632820}

\acmBooktitle{11th International Conference on Digital and Interactive Arts (ARTECH 2023), November 28--30, 2023, Faro, Portugal}

\acmISBN{979-8-4007-0872-5/23/11}

\begin{document}

\title[Pianterly Reality]{Painterly Reality: Enhancing Audience Experience with Paintings through Interactive Art}

\author{Aven Le ZHOU}
\email{aven.le.zhou@gmail.com}
\orcid{0000-0002-8726-6797}
\affiliation{%
  \institution{The Hong Kong University of Science and Technology (Guangzhou)}
  \streetaddress{No.1 Du Xue Rd, Nansha District}
  \city{Guangzhou}
  \state{Guangdong}
  \country{P.R.China}
}

\author{Kang Zhang}
\affiliation{%
  \institution{The Hong Kong University of Science and Technology (Guangzhou)}
  \streetaddress{No.1 Du Xue Rd, Nansha District}
  \city{Guangzhou}
  \state{Guangdong}
  \country{P.R.China}
}

\author{David Yip}
\affiliation{%
  \institution{The Hong Kong University of Science and Technology (Guangzhou)}
  \streetaddress{No.1 Du Xue Rd, Nansha District}
  \city{Guangzhou}
  \state{Guangdong}
  \country{P.R.China}
}

\renewcommand{\shortauthors}{Zhou and Zhang, et al.}

\begin{abstract}
Perceiving paintings entails more than merely engaging the audience's eyes and brains; their perceptions and experiences of a painting can be intricately connected with body movement. This paper proposes an interactive art approach entitled "Painterly Reality" that facilitates the perception and interaction with paintings in a three-dimensional manner. Its objective is to promote bodily engagement with the painting (i.e., embedded body embodiment and its movement and interaction) to enhance the audience's experience, while maintaining its essence.

Unlike two-dimensional interactions, this approach constructs the Painterly Reality by capturing the audience's body embodiment in real-time and embedding into a three-dimensional painterly world derived from a given painting input. Through their body embodiment, the audience can navigate the painterly world and play with the magical realism (i.e., interactive painterly objects), fostering meaningful experiences via interactions. The Painterly Reality is subsequently projected through an Augmented Reality Mirror as a live painting and displayed in front of the audience.

Hence, the audience can gain enhanced experiences through bodily engagement while simultaneously viewing and appreciating the live painting. The paper implements the proposed approach as an interactive artwork, entitled "Everyday Conjunctive," with Fong Tse Ka's painting and installs in a local museum, which successfully enhances audience experience through bodily engagement.

\end{abstract}


\begin{teaserfigure}
  \includegraphics[width=\textwidth]{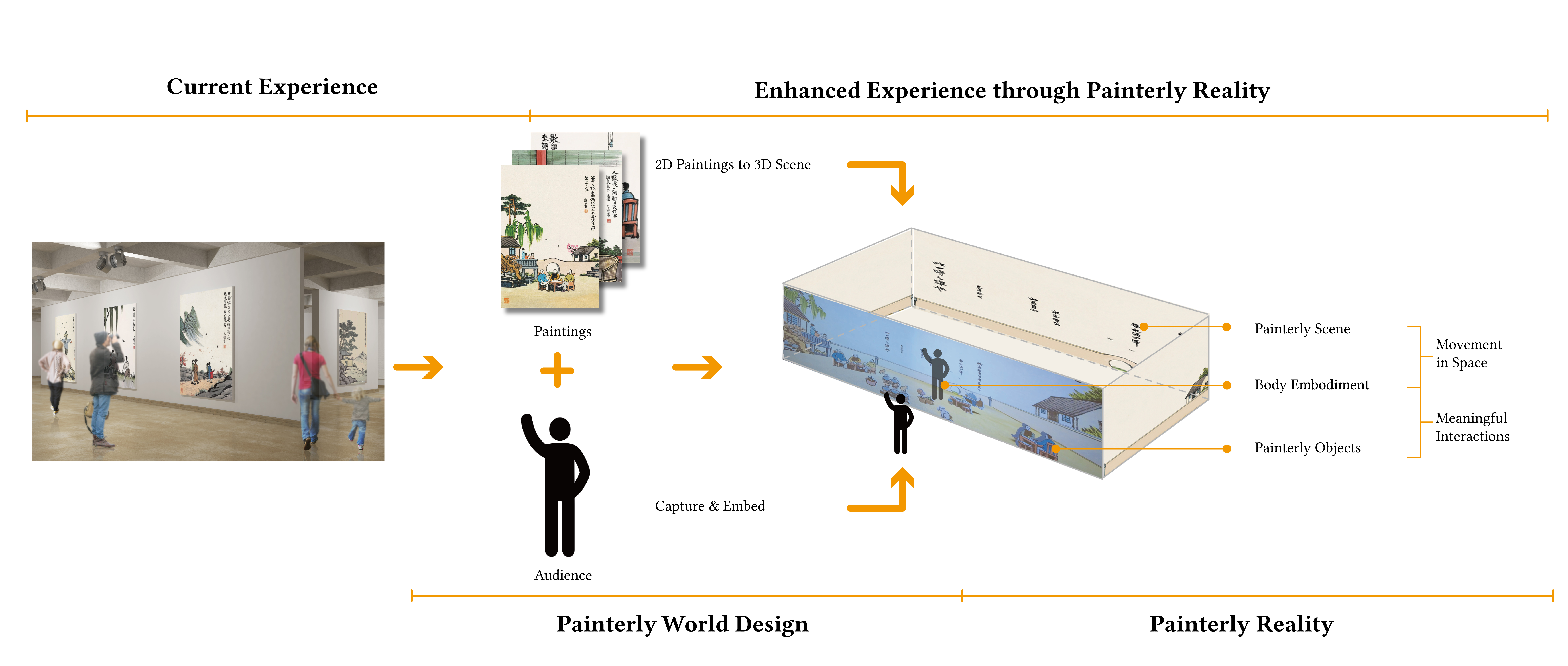}
  \caption{Painterly Reality Design.}
  \Description{A diagram to describe how and why to create painterly reality from painting}
  \label{fig:teaser}
\end{teaserfigure}


\maketitle

\section{Introduction}

As far as we know, paintings are more than 36,000 years old, started with the cave painting of the Altamira in Spain \cite{Pike2012U-series}. Yet this long-established visual art format is being challenged as contemporary audiences are frequently accustomed to engaging with multimedia content such as short-form videos on TikTok that have strong visual appeal \cite{Liu2021Perceived}. Becoming disinterested in paintings and their unique qualities, the audience would be unable to appreciate and engage with the paintings.

\subsection{Audience Experience with Paintings}

Paintings are not experienced in isolation; rather, their perception is profoundly influenced by the viewer's interpretation, response, and the environment in which they are situated \cite{Liu2021Perceived}. As asserted by Mastandrea et al., an aesthetic experience with artwork is characterized by the audience's interactions with the objects and environments, as well as their ensuing reactions \cite{Mastandrea2021Editorial:}. In the realms of pragmatism and phenomenology, bodily engagement has been considered an integral component of art and aesthetic experience theories, with a long-standing tradition \cite{Dewey1934Art,Dewey1980Art, Dufrenne1973Phenomenology, Merleau-Ponty1962Phenomenology}.

The process of perceiving paintings extends beyond the mere utilization of the viewer's eyes and brain; their perceptions and responses are intricately linked to the experience of employing their body \cite{Kühnapfel2023How}. The way the audience moves and attends to their bodies can significantly influence the quality of the experience with artwork \cite{Dewey1948reflex,yip_cinematic_2020}. Nonetheless, this aspect has not garnered sufficient empirical attention \cite{Kühnapfel2023How}. An exploration of the audience's interaction with paintings, encompassing their movement and bodily engagement, remains conspicuously absent and constitutes a research gap that warrants further investigation. 

In contrast to painting that are created by artists and appreciated by the audience, interactive art emphasizes on participatory experiences, such as bodily engagement \cite{Ahmed2018Interaction}. This art form dynamically presents uncertainty and fosters audience engagement with artworks through various interactions and collaborative relationships \cite{Zhang2013Learning}. Supported by emerging technologies \cite{Hu2014Designing, Schraffenberger2012Interaction}, interactive art creates new art experiences that are widely embraced in the contemporary information era \cite{Price2016Exploring}.

\begin{figure}
  \includegraphics[width=0.75\textwidth]{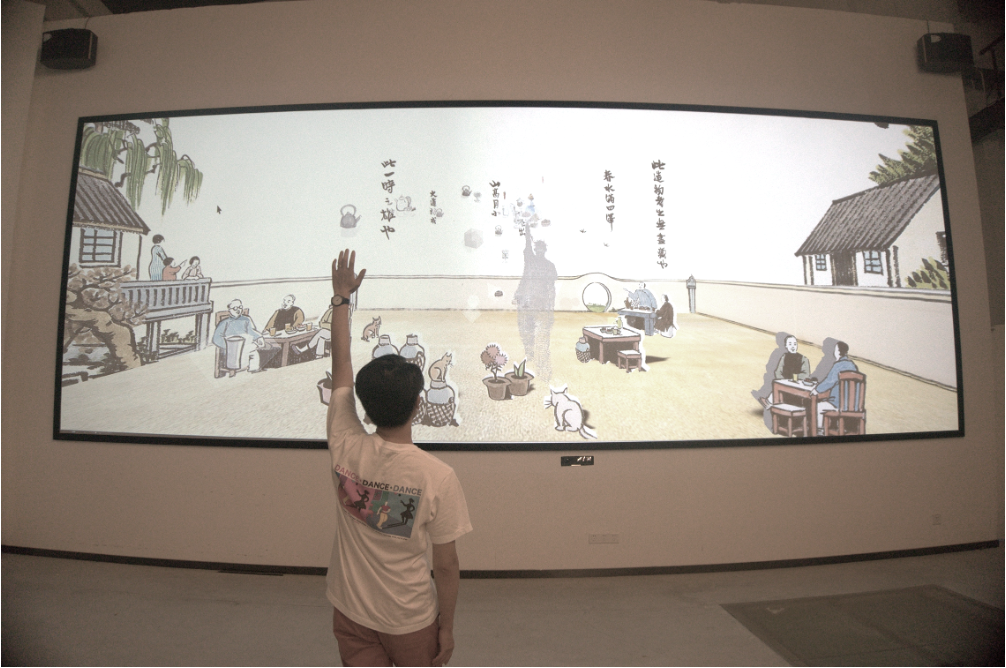}
  \caption{One Participant Interacting with the "Everyday Conjunctive" Installation.}
  \label{fig:EverydayConjunctive}
\end{figure}

\subsection{Painterly Reality}

In Figure \ref*{fig:teaser} and \ref*{fig:EverydayConjunctive} show our proposed 3D interactive art approach, entitled Painterly Reality, involves taking a painting as input and constructing a Painterly World for 3D interaction. Subsequently, we capture and embed the audience's 3D embodiment within the painterly world in real-time. Upon "entering" the Painterly Reality, the audience can navigate the 3D virtual space and interact with the painterly content through bodily engagements. 

The Painterly Reality, embedded with the audience's embodiment, is projected through an Augmented Reality Mirror as a live painting. This enables the audience to gain playful experiences with the painting while simultaneously viewing the live painting with themselves as an integral part. This new approach encourages the audience to engage with an artwork and gain a better understanding of its concept and philosophy in an entertaining manner. Our main contributions include:

\begin{enumerate}
  \item An interactive art approach that aims to encourage and improve audience engagement and experience with paintings. This includes the design of Painterly Reality, which allows the audience's embodiment to interact with painterly contents through an Augmented Reality Mirror. 
  
  \item An interactive installation, entitled "Everyday Conjunctive" shown in Figure \ref*{fig:EverydayConjunctive}, which implements and evaluates the proposed approach with given figurative painting of a well-known artist.
  
\end{enumerate}

We identify the critical challenges in this work as follows: 

\begin{enumerate}
  \item Interpretation of given painting and Painterly Reality design: How to interpret and represent the painting while maintaining its essential focus of aesthetics and philosophy and design the Painterly World? 

  \item Interaction design of the audience with the painterly contents in Painterly Reality: How to connect and engage the audience through meaningful interaction that is intuitive and consistent with the painting's essence?

  \item Implementation: How to present the Painterly Reality through software and hardware system to archive its interaction and design.
\end{enumerate}

\section{Related Work}
Several experimental tasks have been developed to investigate the audience's experience and engagement with paintings. Utilizing Image synthesis algorithms, artistic explorations, e.g., "Mona Lisa Effect" by Xie \cite{XIEMona}, have expanded interactions in portrait paintings. With animated Mona Lisa Painting, it appears as if constantly staring at the audience in response to their movement in front of the painting \cite{XIEMona}. Similarly, Mendoza creates animations from figurative paintings and associate them with kinetic mechanisms, which can be interactively controlled by the audience. It encourages audience engagement through tangible interaction with entertaining mechanics and painterly contents \cite{MendozaMechanical}. However, these examples limit on portrait paintings only and cannot embed the audience's presence into the painting. 

Deep Harmony, employs style transfer-based neural networks, can capture and synthesize the audience into a given painting \cite{Tsai2017Deep}. But the interaction is limited to two-dimensional and lacks real-time functionality. Utilizing Augmented Reality and Deep Learning algorithms, Photo Wake-Up explores the concept of animating a photo or portrait painting, enabling the humanoid painterly content to enter the physical world through Augmented Reality \cite{Weng2019Photo}. This framework can animate humanoid elements in a painting with 3D flexibility, but it lacks interaction or connection with the audience's motion and bodily engagement.

The Flow Room by Jamhoury et al. constructs a 3D virtual space that connects the audience in different locations through network. It captures the audience's embodiment from physical spaces in real-time and presents their telepresence in a unified virtual space \cite{JamhouryFlow}. This new concept of capturing bodily presence, including motion and general bodily engagement, to represent them in a 3D virtual space serves as an inspiration for our approach. While it centers on telepresence within a void space, our endeavor seeks to further extend and enhance the spatial context as a Painterly world, focusing primarily on the audience's experiences with paintings.

\section{Method}

\subsection{Inspiration from Fong Tse Ka}
\epigraph{Art should genuinely reflect people's everyday life and spiritual world to meet art's aesthetics and have eternal vitality.}{\textit{FONG, Tse Ka}}


FONG Tse Ka (FTK) was an influential Chinese painter renowned for his comics-style depictions of daily life and guiding philosophy. He portrays the intricacies of people, everyday life-oriented scenes, and everyday objects in his paintings, adhering to the belief that "art should reflect everyday life and connect with the audience." This delicate alignment with the objectives of interactive art inspires our design of Painterly Reality. 

\begin{figure}
  \includegraphics[width=0.5\textwidth]{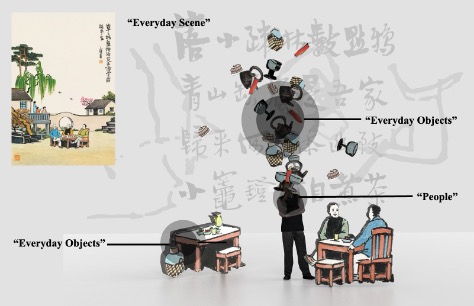}
  \caption{Everyday Scene, Everyday Objects, and People.}
  \label{fig:elements}
\end{figure}

Facing the challenge in interpreting the aesthetic interests and his vital art philosophy, we collaborate with researchers specializing in FTK's art, as well as his descendants, and conduct an empirical analysis of the primary subjects and frequently depicted objects in his paintings and summarize them into three key elements: Everyday Scene, Everyday Objects, and People. We apply these findings as rationale to design Painterly World to capture FTK's essential aesthetics and philosophy.

\subsection{Magical Realism}

Magical Realism is frequently perceived as a fusion of realistic and fantastical elements, resulting in a more comprehensive narrative form than either literary realism or fantasy alone \cite{Wexler2002What}. This approach depicts a realistic worldview while concurrently integrating magical components, thus obscuring the boundaries between fantasy and reality \cite{Bowers2004Magic}. Although magical elements are present, a substantial amount of realistic detail is incorporated, utilizing these magical aspects to convey a deeper understanding of reality \cite{Cortés1992Dictionary, Wexler2002What}. 

We consider the Painterly World constructed from FTK's "three elements" as the "mundane" setting, representing the first layer of realism and functioning as the real-world backdrop. Subsequently, we introduce fantasy and magical elements to this "realistic" view by incorporating interaction with the painterly objects as the supernatural effects within Painterly Reality. 

\subsection{Design Rationale}

\subsubsection{Painterly World Design}

FTK's paintings, characterized by their moderately 3D depth, inspire us to design the Painterly Reality in 3D with the three key elements (i.e., Everyday Scene, Everyday Objects, and People). We propose a 3D Painterly World should include: 

\begin{enumerate}
\item A 3D Painterly Scene provides a spatial foundation for seamlessly mapping the audience's movement in the physical space and interactions to Painterly World. 
\item 3D Painterly Objects derived from the paintings and positioned in Everyday Scene, serve to augment the perception of depth, and contribute to the realism of the painterly world. 
\item The audience's Body Embodiment, as the People in FTK's painting, embedded within the painterly world. 
\end{enumerate}

Despite the spatial arrangements and design guidelines, achieving audience's participatory experiences remains a challenge. Analogous to an empty stage awaiting a performance, the remaining significant challenge is to connect the audience and their body embodiment to the Painterly World in spatial context.

\subsubsection{The Audience's Body Embodiment and Movements in Space}

The audience and their body embodiment can move left or right, approach closer or further, within the Painterly World, allows them to grasp the painting's philosophical and aesthetic essence through their participations. The audience's embodiment directly contributes to the real-time construction of the Painterly World, serving as an essential element of FTK's paintings, which enables them to witness themselves becoming a part of the Painterly Reality and facilitates an immersive experience. In this way, the art "actualizes its reality when connected with the audience."

\subsubsection{Meaningful Interaction}

We incorporate the concept of Magical Realism and its premise of adding magical elements and fantasy to reality \cite{Wexler2002What}, so to devise intuitive and meaningful interactions that align with the painting's essence. In contrast to particle systems employed for abstract visuals, commonly found in media art projects, we designate the audience's virtual embodiment as the particle emitter and FTK's Everyday Objects as particles. This approach ensures that the interaction is meaningful and remains true to the style and content of the painting.

\subsection{Augmented Reality Mirror}

The Augmented Reality Mirror (ARM) is a specular interface that virtually alters the reflective scene in the mirror using augmented reality \cite{Azuma1997survey}. By incorporating camera positioning within a spatial context, it integrates users as part of an augmented scenario \cite{Azuma1997survey}. ARM combines real and virtual objects registered in three dimensions and facilitates real-time interaction \cite{Portalés2018Interacting, Azuma1997survey}. To construct and present the Painterly Reality in real-time, we propose the ARM system as a hardware solution for Painterly Reality.

Through ARM, a camera captures the audience's embodiments, integrating them into the Painterly Reality and facilitating real-time and meaningful interactions with the magical elements. In the fundamental layer of realism, the system functions as a reflective interface, and the augmented mirror combines real and virtual objects in 3D. The layer of magical realism engages the audience with the painting through bodily interactions that remain true to the artist's style and philosophy. 

Simultaneously, the generated Painterly Reality is projected onto a screen as a live painting (as shown in Figure \ref*{fig:EverydayConjunctive}), enabling the audience to interact with it three-dimensionally while appreciating the generated content and aesthetics, with themselves fully immersed. In this manner, the audience assumes the dual roles of viewer and co-creator of the painting. This cyclical process further amplifies the audience's engagement with the artwork.

\begin{figure}
  \includegraphics[width=0.75\textwidth]{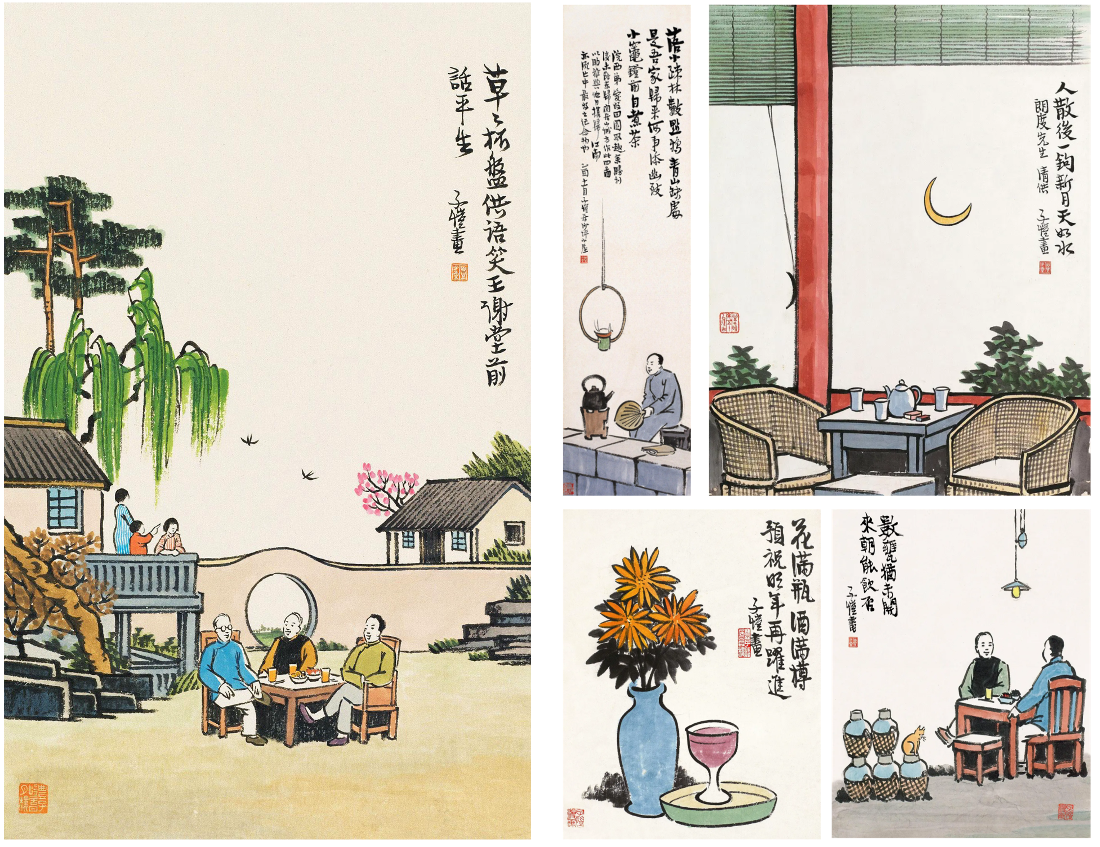}
  \caption{(left) Painterly Scene Design Reference (right) Everyday Objects in FTK's paintings.}
  \label{fig:refimg}
\end{figure}

\section{Experiment: Everyday Conjunctive}

We delve the proposed Painterly Reality approach with FTK's painting as an interactive installation, entitled "Everyday Conjunctive" and display in a local exhibition hall of FTK. We select a representative figurative painting by the artist, as depicted in Figure \ref*{fig:refimg} (left), and incorporate a capturing module to embed the audience's embodiment and a particle-based method to facilitate meaningful interactions, ultimately show the painterly reality on a projection screen.

\begin{figure}
  \centering
  \begin{subfigure}[a]{\textwidth}
    \includegraphics[width=\textwidth]{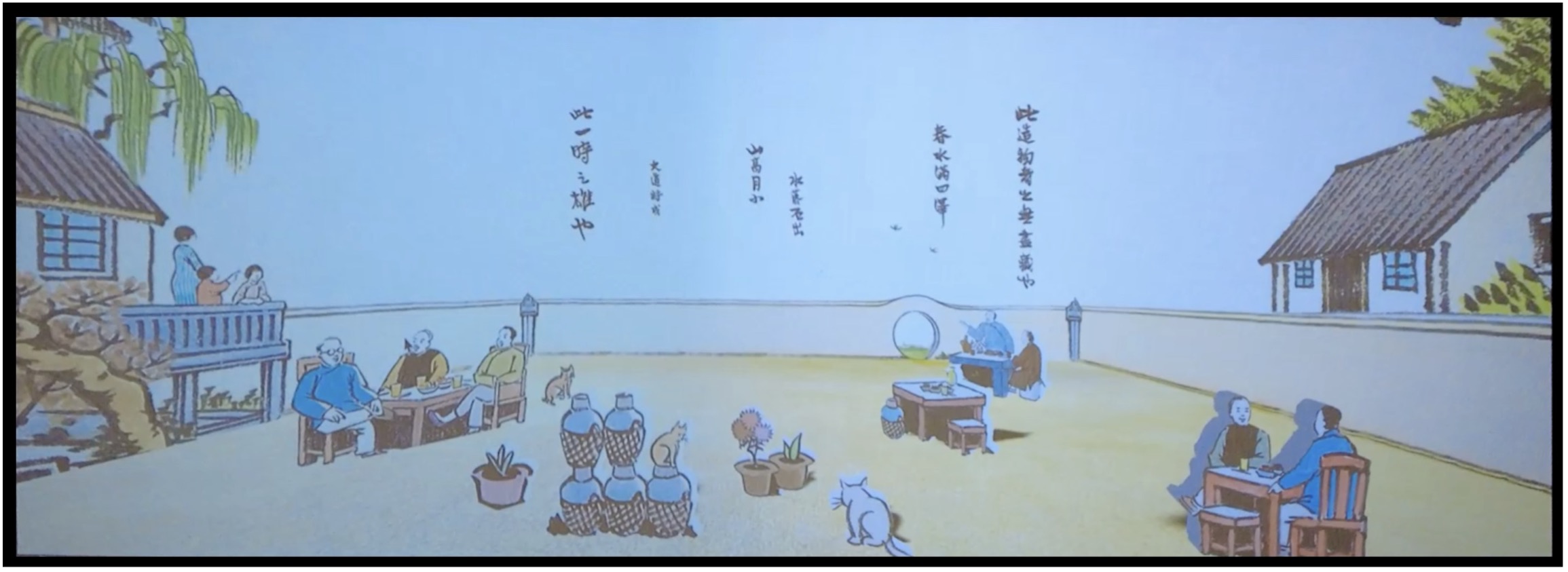}
    \caption{3D Painterly Scene}
  \end{subfigure}

  \begin{subfigure}[a]{\textwidth}
    \includegraphics[width=\textwidth]{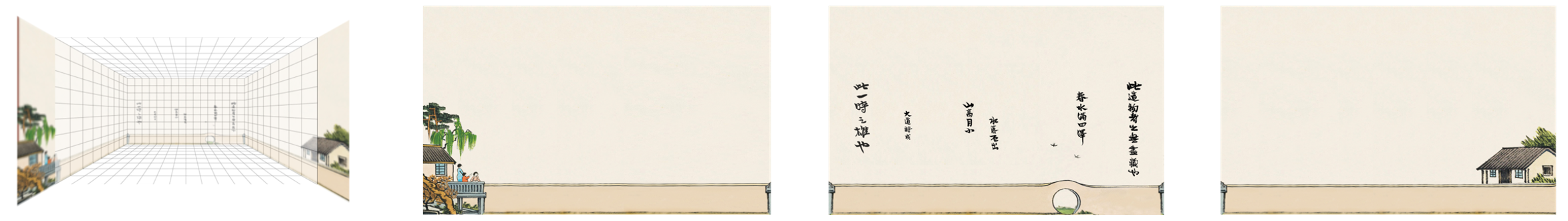}
    \caption{Three-Sides "Facade"}
  \end{subfigure}

  \caption{(a) 3D Painterly Scene from (b) Three-Sides "Facade."}
  \label{fig:facade}
\end{figure}

\subsection{Design the Painterly World}

Figure \ref*{fig:facade} (a) shows we construct a 3D courtyard space, i.e., the Painterly Scene, encompassed by three-sides "facade." Figure \ref*{fig:facade} (b) showcases the image collage composed of visual elements derived from FTK's painting. These "facades" are assembled as a virtual room with the 3D depth that enables the audience's virtual embodiment to navigate the space. 

Furthermore, we curate a collection of "Painterly Objects" i.e., the frequently depicted elements in FTK's paintings (see examples in Figure \ref*{fig:refimg} (right)). We identify them and design them into 3D objects with FTK's researchers and descendants. Subsequently, we strategically place some of them within the Painterly World to embody FTK's aesthetic interests and enhance the design. Consequently, we transform the 2D painting into a 3D Painterly World that allows the audience's body embodiment to engagement with.

\begin{algorithm}
  \caption{Capture 3D Point Cloud}\label{alg:capture}
  \begin{algorithmic}[1]
  \Require $\text{depthImage, depthList}$
  \Ensure $\text{pointCloud}$
  \State $\text{depthThreshold} \leftarrow \text{some value}$
  \For{$y = 0$ to $\text{height of depthImage} - 1$}
  \For{$x = 0$ to $\text{width of depthImage} - 1$}
  \State $\text{depthValue} \leftarrow \text{depthList}[y \times \text{width of depthImage} + x]$
  \If{$\text{depthValue} \in [\text{depthThreshold} - \epsilon, \text{depthThreshold} + \epsilon]$}
  \State $\text{pointCloud} \leftarrow \text{pointCloud} \cup \{(x, y, \text{depthValue})\}$
  \EndIf
  \EndFor
  \EndFor
  \State \Return $\text{pointCloud}$
  \end{algorithmic}
  \end{algorithm}
  
\subsection{Embed the Audience's Body Embodiment}

We employ a custom algorithm with the Microsoft Kinect to capture the audience's entire body in 3D point cloud in real-time and serve as their embodiments. This method first obtains depth map images from the dual camera, using the depth information to eliminate objects in the background and construct the 3D point cloud.  It can then be easily incorporate to painterly world allows the participant to navigate within. The implementation detail of how to capture 3D point cloud is shown as pseudo-codes, which can seamlessly integrate with the interaction design and custom algorithm explained in next subsection.

\subsection{Add Magical Realism and Interaction}

We conceive a new interactive design that guides the audience's interaction with the Painterly Objects. As the audience enters the installation, their heads or raised hands become particle emitters that continuously generate and cast out the Painterly Objects as particles. When the audience dances, jumps, or moves around, the constantly generated Painterly Objects bounce out, fly upwards, and ultimately fall following the particle behavior design. 

The custom particle system simulates various physical behaviors, such as movement, bounce, and gravity. A particle originates from a given position with an initial speed and lifespan, flying upwards with increasing speed and decreasing lifespan over time. Gravity affects the speed in the vertical direction, causing the particle to fall to the ground. Upon hitting the ground, the vertical speed "flips" to create a bouncing effect. 

Our custom particle (generator and) emitter algorithm creates and updates a list of particles through random sampling from the Painterly Objects list. It adds particles one by one from the beginning until reaching the maximum amount in the particle array. Particles are automatically removed from the list upon reaching the end of their lifespan, and new particles are created. In this manner, the algorithm continuously generates new particles and updates the particle array. We present the mechanism as the following pseudo-codes in details.

  \begin{algorithm}
    \caption{Particle Emitter}\label{alg:emit}
    \begin{algorithmic}[1]
    \Require $\text{particleList}$
    \Ensure $\text{particleNum} = 50, \text{Buffer} = 50$
    \ForAll{$\text{point} \in \text{pointCloud}$}
    \If{$\text{length}(\text{particleList}) < \text{particleNum} \And \text{Buffer} = 0$}
    \State $\text{particleList} \leftarrow \text{particleList} \cup \{\text{new particle at point position}\}$
    \State $\text{Buffer} \leftarrow 50$
    \Else
    \State $\text{Buffer} \leftarrow \text{Buffer} - 1$
    \EndIf
    \EndFor
    \ForAll{$\text{particle} \in \text{particleList}$}
    \If{$\text{lifespan}(\text{particle}) < 0$}
    \State $\text{particleList} \leftarrow \text{particleList} \setminus \{\text{particle}\}$
    \EndIf
    \EndFor
    \end{algorithmic}
    \end{algorithm}

\begin{figure}
  \includegraphics[width=10cm]{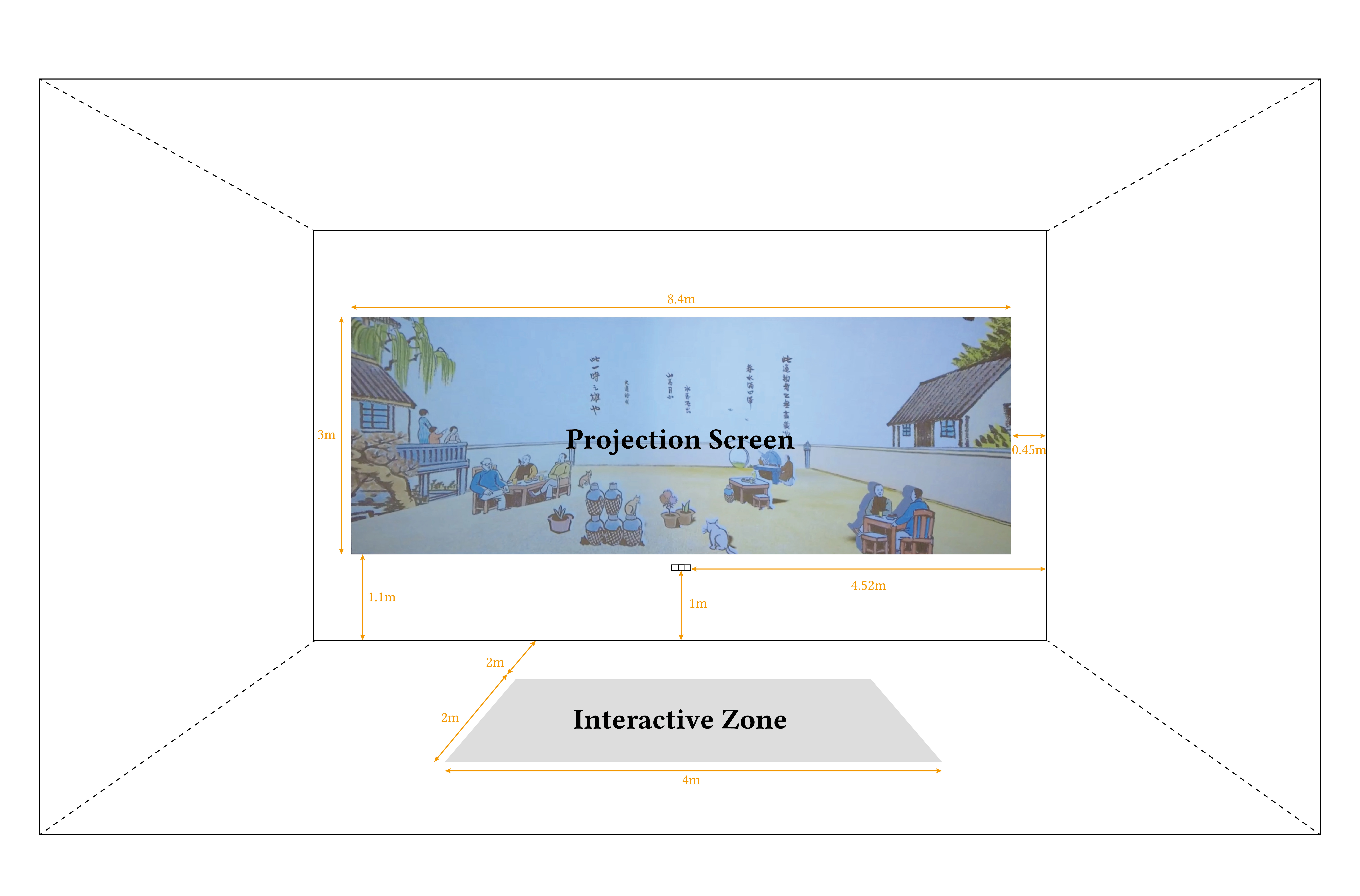}
  \caption{Installation Dimensions \& Spatial Setup.}
  \label{fig:dimension}
\end{figure}
  
\subsection{Project to ARM and Deployment}

We deploy "Everyday Conjunctive" with a large projection screen in a local museum of FTK as illustrated in Figure Figure \ref*{fig:EverydayConjunctive}, the participant's embodiment appears in a scale identical to the actual human. When the participant "enters," their body embodiment performs correspondingly on the screen like a mirror, with the augmentations. Figure \ref*{fig:dimension} shows the installation dimensions in detail.

\section{Evaluation}

To evaluate the efficacy of our proposed approach and implementation, we observe on audience perceptions and reactions to Everyday Conjunctive and collect data through on-site surveys and interviews. The evaluation was conducted during the first two days after the installation open to the public. We initially document the audience's behavior and subsequently approach randomly selected individuals for survey and interview. 

The audience engage with the installation as part of their museum visit. Before interacting with the installation and participating in the surveys and interviews, we refrain from interacting with the audience or providing any verbal explanations, such as the interaction design or project aims, to allow for independent exploration. An artwork plaque displaying the project title, year, material, and artist names is the only accompanying information provided on-site.

\subsection{Observation Summary}

We observe that the audience can readily transfer their experiences and understanding of how mirrors work to Everyday Conjunctive. With minimal explanation about the installation or interaction, they can intuitively comprehend the setup and promptly begin to play. Once immersed in the 3D Painterly Reality, the audience can quickly discover the interaction. This primarily occurs because they can see themselves within the mirror, as if they were walking into an actual "space." 

Upon discovering the playful interaction, they swiftly commence their exploration. A prevalent behavior among numerous participants involves jumping as high as possible, waving their hands rapidly, or running around with exaggerated gestures. These actions seemingly aim to assess whether the installation operates as anticipated and whether it can withstand their "pressure test" and consistently performing well with "extreme" inputs.

\subsection{Survey and Interview}

Overall, 16 individuals (out of 28 audiences who interacted with the exhibit during the two-day observation) agree to participate in the survey and interview. Their play with the exhibit and engagement ranges from 11 minutes to 24 minutes with an average of 15.2 minutes. After they finish playing with the exhibit, we approach them for this qualitative evaluation. Some answers are collected through multiple-choice questions while other data are acquired through participants' verbal descriptions during the interview. We prepare a series of questions as below. 

\begin{enumerate}
  \item The participants' basic information and prior exposure to FTK's artwork and philosophy, as well as experience of (any kind of) Interactive Art, Interactive Screens, and ARMs.
  \item The participants' feedback and evaluation on the design and interaction including:
  \begin{enumerate}
    \item  Is the interaction working intuitively or difficult-to-understand? How have you figured it out?
    \item {If participant knows FTK}: Does the exhibit capture FTK's aesthetics?  
    \item Is the exhibit a validation or enhancement of the expression of FTK's aesthetics and philosophy?
    \item Does the interactive exhibit create connection and encouraging experience with FTK? (Optional) How?
    \item Is it improving the engagement with painting comparing to traditional showcase? (Optional) How?
    \item {Otherwise}: Do you capture any FTK's aesthetics from the interactive exhibit, what are they?
    \end{enumerate}

  \item The participants' suggestions about the art installation. 
  \end{enumerate}

Participants (P1-P16) are mostly adults aged from 23 to 59 except for one 12 years-old child (p7), 9 male and 7 female, 11 of them participate in the first day and 5 of them in the second day. The project is installed in FTK's hometown in China Mainland and most of the audience are Chinese. 6 participants (P3, P5, P6, P12-P14) are locals and the other 11 are tourists, thus they mostly know about FTK and his art. Only 2 participants (P4, P16) are not familiar with FTK's art and background, the others are familiar (9 out of 14) or very familiar (5 out of 14) with FTK. 13 participants (P1-P5, P7-P12, P15-P16) have previous experiences with interactive art, as well as experience with interactive screens except for P16, but none of them experienced ARMs before. 

All 16 participants find the interaction highly intuitive except P1 and P4 who find it "more or less intuitive" to understand. It matches our observation onsite; the audience pick up the interaction and start their exploration very quickly once they see themselves on the screen. 10 out of 16 participants describe their understanding of the interaction as intuitive without further thoughts proves our observation. 4 out of 16 participants imitate the interaction from watching previous audience. 

P4 and P16 are not familiar with FTK, P4 find the exhibit significantly helps understanding FTK's art. All 14 participants that are (very) familiar with FTK agree the exhibit at least sufficiently capture the artist's essential style and enhance the expression of FTK's aesthetics and philosophy. 6 out of 14 participants (P1-P3, P8-P10) evaluate highly of the design and interaction (with all three highly-agree answer to questions b – e.)  

P7 describe the exhibit encourages his interest of FTK because the particle - Everyday Objects "like candy rain, they bounce around made me want to catch them and see what exactly they are" and "never realize FTK is so interesting" and went back to play the interaction again after the interview. P3 are very interested in the hardware and software system and think "it is much more interesting than the traditional exhibition format and kid will love it" and "it should be applied to more traditional cultural painting to bring people's attention, and museums should adapt to such kind new technologies otherwise they have no future." 

When comparing this exhibit with paintings, P10 comments it as "not comparable" since "interactive art has its great advantage in audience engagement" and "to take advantage of the nature of Interactive Art and apply on painting will surely succeed the traditional artwork in terms of audience engagement."

Four participants (P1, P3, P10, P11) suggest at least one comment (part 4 of the interview), and P1 likes the background music and praises about the sound effect which are off the focus of this paper. P1 also suggests having a "screenshot" and sharing function since he "want the painting with himself inside the FTK painting he participated and helped to create." P3 and P11 both wish it could have more scenarios to experience with different paintings. P11 also suggests on haptic feedback which we discuss in Section 6. P10 points out about the distortion of the captured embodiment when getting too close to the screen, which is due to the dual-camera's effect of foreshortening.

\begin{figure}
  \includegraphics[width=\textwidth]{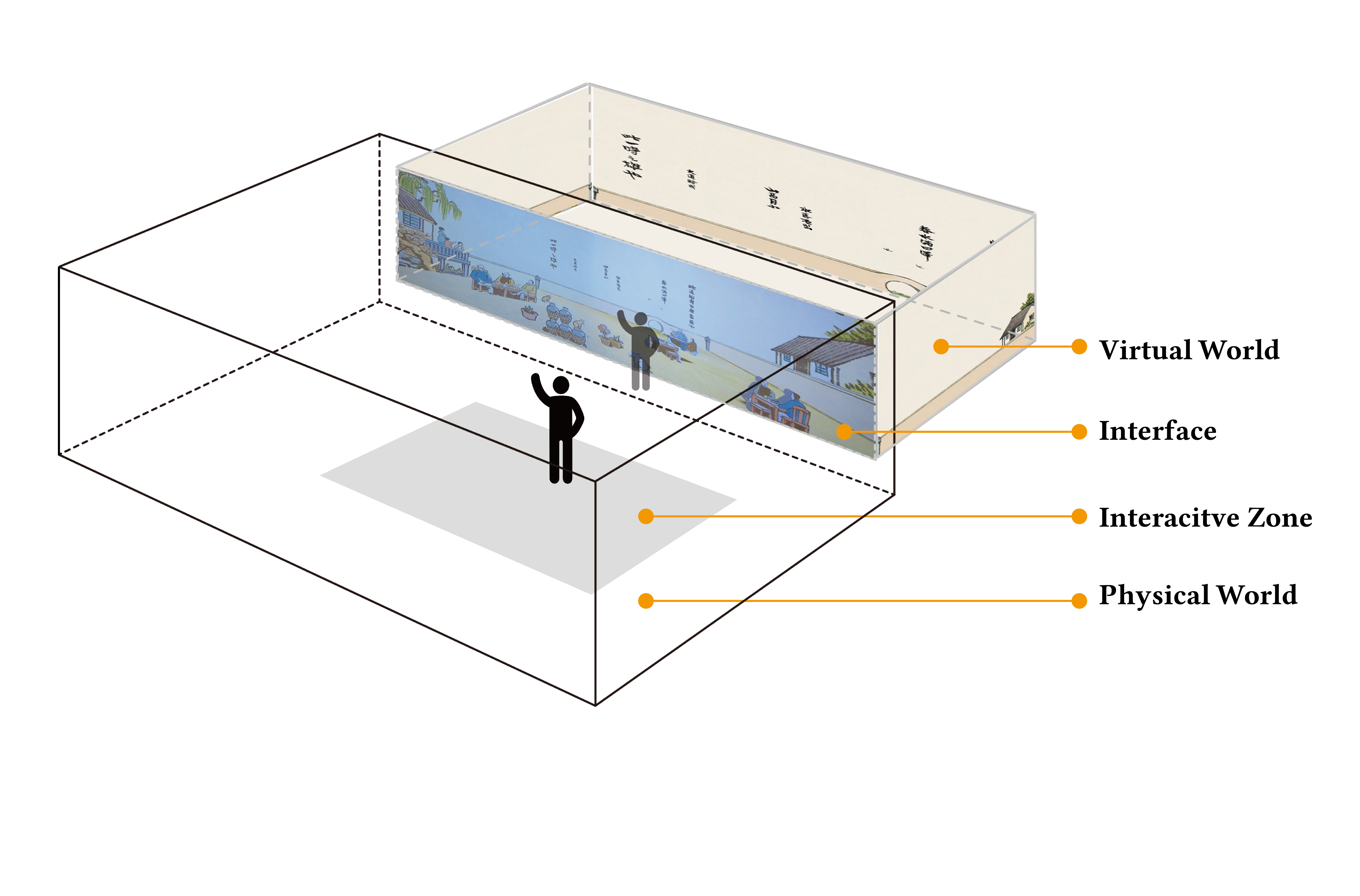}
  \caption{Augmented Reality Mirror - Physical vs. Virtual World \& the Interface In-between.}
  \label{fig:inbetween}
\end{figure}

\section{Discussion}

Nearly all participants concur that the design and interaction function intuitively, enabling them to easily comprehend and engage with the exhibit. The proposed approach has effectively captured the essential aesthetics and philosophy of the artist's "everydayness" concept, successfully encouraging, and enhancing engagement with FTK's paintings for all the audience.

By creating a Painterly Reality from given paintings, we bridge the physical world with the virtual one, allowing the audience's embodiments to "enter" FTK's painting (see Figure \ref*{fig:inbetween}). The interaction with Painterly Objects introduces fantasy and magical elements to FTK's "realistic" realism. Both the interaction and design of Painterly Reality are meaningful and adhere to the style and content of the artist's painting.

The audience participates in the exhibit as both co-creator and viewer of a painting. Painterly Reality serves as a platform for the audience to engage with the painterly content and act as a live visual element in the three-dimensional painting. Simultaneously, the audience engages with the exhibit as a viewer, appreciating the generated painting displayed through the augmented reality mirror. The process of creating and viewing the generated painting via Painterly Reality becomes cyclical, cumulatively enhancing the audience's engagement with the painting.

\section{Limitations and Future Work}

A limitation of the current study lies in the small sample size of interview participants in the evaluation. We plan to conduct further interviews with more participants and gather additional data for subsequent analysis. Frequently mentioned requests include incorporating more scenarios with different paintings and exploring the potential application to other similar paintings. 

The proposed approach can be easily adapted to other artists' paintings by creating more painterly scenes, while maintaining the same interaction and design to effectively scale up the experience and further encourage engagement with the paintings. Once a three-dimensional scene of a given painting is constructed, the audience can participate as part of the live painting through their embodiments. This multi-level realism transforms real-time interaction and participation into a "painting in which the participant contributed to its creation."

A thought-provoking question and suggestion from a participant offer an intriguing perspective. The participant stated, "I enjoy the illusion that I am inside FTK's painting and interacting with the objects, but I cannot physically feel them, and it breaks the illusion and immersion immediately." While we have connected the physical world and the virtual Painterly Reality with sufficiently eligible interactions and participatory experiences, these occurrences are solely visual. The swiftly formed immersion and enjoyment dissipate quickly due to the absence of physical and haptic responses. In future experiments, we are considering the implementation of haptic feedback, potentially utilizing VR controllers, for the Painterly Objects interaction.

\begin{acks}
  We thank Yida He, Fuqi Xie and Molly He for their assistance during the implementation of "Everyday Conjunctive," especially the visual material design and onsite installation. We thank Xufan Lin's help on design of Figure 1, 6 \& 7. 
\end{acks}
  
\bibliographystyle{ACM-Reference-Format}

\bibliography{Artech_Painterly_Reality.bib} 
\end{document}